\documentclass[twocolumn,showpacs,preprintnumbers,amsmath,amssymb]{revtex4} 
\usepackage{graphicx} \usepackage{color} 
\usepackage{bm} 
\begin{document} 
\newcommand{\rf}[1]{(\ref{#1})}
\newcommand{\bfomega}{ \mbox{\boldmath{$\omega$}}}
\title{Fluorescent contacts measure the coordination number\\and entropy of a 3D jammed emulsion packing.}  
\author{ J. Bruji\'c \footnote{Present address: Biological Sciences, Columbia University, New York, NY 10027 }}
\affiliation{ Polymers and Colloids Group, Cavendish Laboratory, University of Cambridge, Madingley
Road, Cambridge CB3 0HE, UK}
\author{G. Marty, C. Song, C. Briscoe, H. A. Makse}
\affiliation{ Levich Institute and Physics Department, City College of New York, New York, NY 10031 }
\begin{abstract}
Jammed matter is by definition impenetrable to light,
rendering the characterization of the 3D geometry difficult.
Confocal microscopy of a dyed, refractive index matched emulsion
nevertheless allows one to image the jammed system.
Here we explain the origin of the mechanism of enhanced fluorescence at the
contacts of jammed emulsion droplets in terms of a blue-shifted
fluorescence emission band due to the change in the polarity of
the interfacial environment.
We then use this information to determine the contact network in the
emulsion, which models a frictionless jammed system.
This enables the  experimental determination of the 
average coordination number, $\langle Z \rangle$ giving 
the theoretically predicted value of 
$\langle Z \rangle \approx 6$. Furthermore, the method
enables the experimental measurement of the entropy of the packing
from the network of contacts.
\end{abstract}
\maketitle

Pouring sugar into a cup is the simplest example of a fluid to solid transition which takes place solely because of an increase in particle density, identified as the jamming transition \cite{liu}. In other words, if the particle density is high enough to support an external weight without the constituent grains starting to flow and rearrange, the material has the properties of a solid and is known to be at the isostatic limit. This state arises when the particles have enough contacts between them such that all the forces balance according to Newton's equations \cite{ed1}. The average number of contacts per particle, $\langle Z \rangle$, known as the coordination number, is therefore the key parameter that determines the mechanical properties of granular materials. Theory predicts that a system of smooth frictionless particles has a minimal average number of contacts for mechanical stability $\langle Z \rangle =2D$ \cite{alex}, where $D$ is the dimension of the system.\\
The problem with jammed matter, from sand piles to mayonnaise, is
that it is difficult to take a look inside the particulate packing
to measure $Z$. Indeed, in the old days Mason, a postgraduate
student of Bernal, took on the task of shaking glass balls in a
sack and "freezing" the resulting configuration by pouring wax
over the whole system. He would then carefully take the packing
apart, ball by ball, noting the positions of contacts for each
particle \cite{bernal}. Since this labor-intensive method patented
half a century ago, still used in recent studies \cite{donev}, other
groups have extracted data at the level of the constituent
particles using X-ray tomography \cite{richard}. However, neither
method could directly determine the contacts in order to be sure
whether the particles were touching or just very close together.
Moreover, these methods only provide insight into large particles
and do not lend themselves to the exploration of jammed matter on
the colloidal length scale. Given the recent boom of theories
unifying the concepts in granular materials and glasses
\cite{coniglio}, the characterisation of all jammed matter is
crucial for the advancement of this field.\\
In this paper, we explain a different method which 
directly highlights the contacts between droplets in a jammed emulsion.
While previous experiments
were based on the visualization of the areas of droplet deformation 
\cite{brujic}, here we explain the origin of the enhanced fluorescence 
at the contacts in terms of the polarity of the
environment of the dye. This information allows the reconstruction of
the contact network for any external pressure of the system, in
particular that of the low pressure isostatic limit.
Using a combination of UV-Vis spectroscopy and fluorescence microscopy
we present a measurement of $\langle Z \rangle$ for a frictionless jammed
material and a measurement of the entropy in jammed systems, all this at the
colloidal lengthscale.

\begin{center}
\begin{figure}[h!]
\includegraphics[width=55mm]{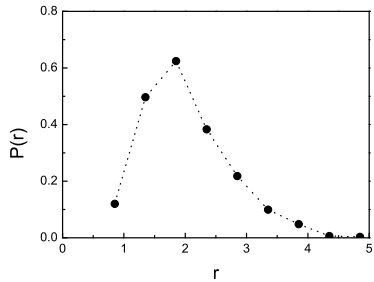}
\caption{Radius distribution of the droplets.}
\label{Pr}
\end{figure}
\end{center}

{\it{Emulsion system}} - The model system used in this work to study jammed systems is an emulsion. It consists of silicone oil droplets in water, of the order of 4$\mu$m in size. The radius distribution is shown on Fig.\ref{Pr}. This emulsion system is stabilized by Sodium Dodecyl Sulfate (SDS) surfactant, at 10mM, i.e. below the Critical Micellar Concentration (CMC) thus avoiding depletion attraction between droplets \cite{mason}. It can then serve as a model for a granular material composed of soft spheres wich can be considered frictionless due to the lubrication by the continuous phase fluid. 
In order to enable observation in a fluorescence scanning confocal microscope (LSCM), glycerol is added to the continuous phase in order to obtain a transparent sample by mean of index matching, and the silicone oil was fluorescently dyed with nile red \cite{brujic}.
\begin{center}
\begin{figure}
\includegraphics[width=70mm]{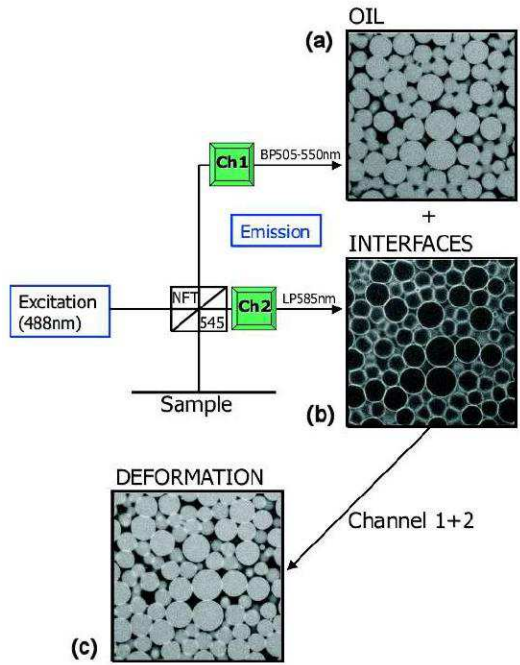}
\caption{Confocal microscope images obtained by splitting the fluorescence using optical filters, (a) BP505-550nm and (b) LP585nm, and their subsequent superposition (c) providing the contrast necessary for image analysis. Note that contacts are highlighted.}
\label{setup}
\end{figure}
\end{center}
%\vspace{-1.5cm}
{\it{The method}} - As will be explained later, the activity of nile red depends on its environment. Then the fluorescence of different regions of the sample might be observed in different spectral domains. The method thus consists in exciting the dye at a wavelength of 488nm and split the emitted signal into two channels, CH1 and CH2, as shown in Fig. \ref{setup}. CH1 is equiped with a band pass filter from 505nm to 550nm, and gives the image of Fig. \ref{setup}a. The signal in CH2 goes through a long pass filter above 585nm and the resulting image is shown in Fig. \ref{setup}b. As one can see, this method enables to separate the contributions of the droplets cores and their interfaces in the fluorescence signal, while the continuous phase gives no significant emission. It is then possible to reconstruct an image in which intensities of both droplets and interfaces are equalised, as shown in Fig. \ref{setup}c. On this image, one notes that areas of contacts between droplets are highlighted. It corresponds to the situation where {\it{i)}} the two interfaces are closer than the microscope resolution so that their activities add as they are detected by the same voxel and {\it{ii)}} the deformation  zone resulting from the contact between two droplets has an area bigger than the resolution of the microscope so that at least one voxel is highlighted. The second condition ensure that we are actually looking at real contacts, and not at very close droplets. Given the experimental conditions of this study, this condition is achieved for pressure greater than $0.1 \sigma /r$, where $\sigma$ is the surface tension of the interface and $r$ the radius of the droplets, which was realized by centrifugation of the emulsions. This method then enables a complete study of the contact network in a 3-dimensionnal sample as will be illustrated after having presented an investigation to understand the physical processes underlying these effects.
\begin{center}
\begin{figure}[h!]
\includegraphics[width=60mm]{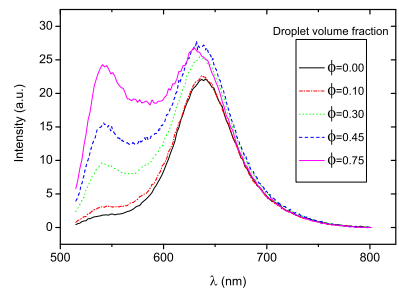}
\label{emspec}
\caption{Emulsion emission spectra for different volume fractions from 0 to o.75.}
\end{figure}
\end{center}
%\vspace{-1cm}

\begin{center}
\begin{figure}[h!]
\includegraphics[width=80mm]{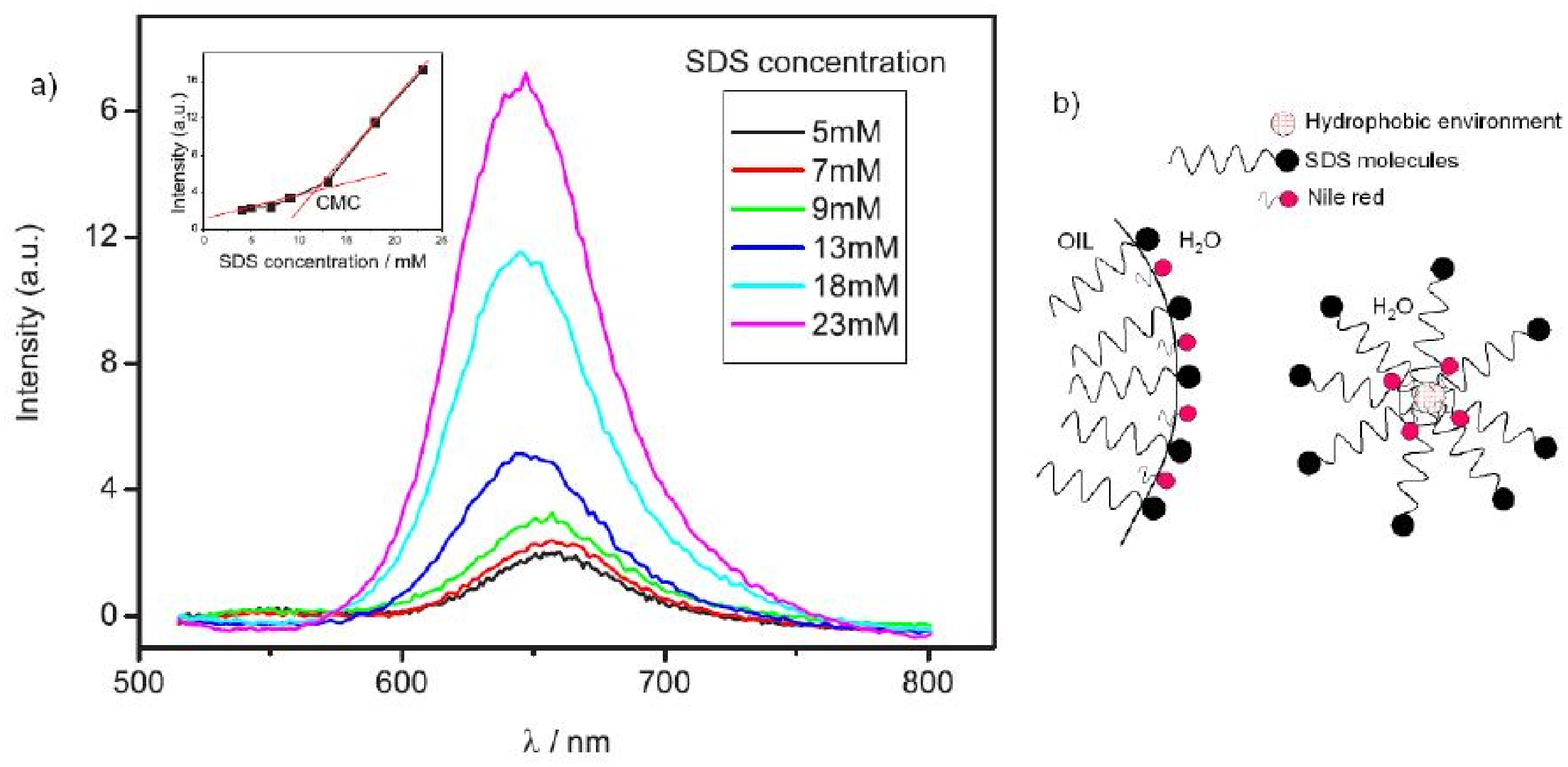}
\caption{(a) Emission spectra of the aqueous phase at a constant nile red concentration, but increasing SDS concentration. The maximum intensity is plotted versus the SDS concentration in the inset, showing a change of behavior at a concentration which correponds to the CMC. (b) Schematical representation of nile red insertion at interfaces (right) and in micelles (left).}
\label{emspec2}
\end{figure}
\end{center}

\begin{center}
\begin{figure}[h!]
\includegraphics[width=80mm]{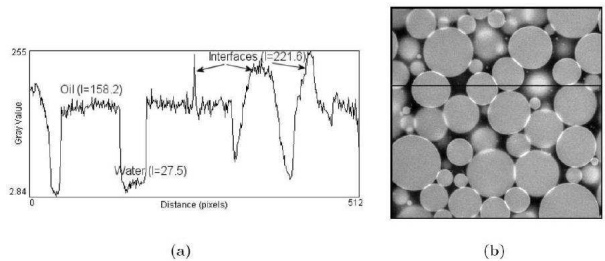}
\caption{(a) Intensity profile along the black line of image (b). The ratio of intensity between interfaces and the continuous phase is $221.6/27.5 \simeq 8$.}
\label{voxel}
\end{figure}
\end{center}

{\it{Environment-sensitive fluorescence of nile red}} - We now describe UV-visible spectrometer experiments showing that the fluorescence of nile red depends on its environment, then allowing to filter the different contributions in confocal microscope images, as has been discussed above. To investigate the different contributions, it is interesting to consider the emission spectra of the emulsion system varying the proportion of the different species, i.e. aqueous phase, droplets and interfaces. A convenient way to vary this parameter is to make emulsions at different volume fractions. The results are presented on Fig. \ref{emspec}, for volume fraction from $\phi=0$ (purely aqueous phase) to $\phi=0.75$. The peak corresponding to the continuous phase can be identified easily thanks to the spectra at $\phi=0$, and is located around $\lambda=640$nm. When the volume fraction is increased, two effects can be observed. First, a peak at shorter wavelengths appears and grows linearly with the volume fraction. Second, the peak at $\lambda=640$nm increases, whereas the proportion of the aqueous phase decreases. The amount of this augmentation is about $22\%$ between $\phi=0$ and $\phi=0.75$. It is then clear that droplets contribute to the fluorescence at two wavelength, namely $\lambda=545$nm and $\lambda=640$nm. According to the effect observed in the confocal images and reported above, as well as to emission spectra measurements of nile red in pure oil which gives a peak around $545$nm, we deduce that the contribution at $\lambda=545$nm comes from the droplets cores and that the one at $\lambda=640$nm comes from the interfaces.\\
Consequently, there are two questions which are to be addressed : {\it{i)}} why is the signal of the interface similar to the one of the continuous phase and not to the one of the droplets ? and {\it{ii)}} why does interfaces appear much brighter than the continuous phase ?\\
To answer the first question one can study the emission spectra of the aqueous phase for increasing SDS concentrations, as shown in Fig. \ref{emspec2}a. The clear enhancement, particularly above the CMC, suggests that micelles play a crucial role in the solubilization of the dye. The answer can then beformulated as schematically represented on Fig. \ref{emspec2}b \cite{wagner}: the hydrophobic parts of nile red molecules might insert themselves in the central regions of micelles, away from water, while their other end is in contact with the polar solvent, just as it is the case at the interfaces. On the contrary, nile red molecules in the cores of droplets are not at all in contact with water, having a blue shifted emission since their environment is much less polar. \\
The fact that interfaces appear brighter than the continuous phase (even though they both emit in the same spectral range) can be understood by a simple semi-quantitative analysis, if one asumes that the peak intensity in the emission spectra is proportionnal to the quantity of dye. In this hypothesis, the peak intensity of the pure aqueous phase is $P_0 \propto c_w$, where $c_w$ is the concentration of dye in the aqueous phase. The same peak at $\phi=0.75$ has an intensity $P\propto c_w \phi _w+c_i \phi _i$, where $c_w$, $\phi _w$, $c_i$ and $\phi _i$ refers respectively to the dye concentration and the volume fraction of the aqueous phase and the dye concentration and the volume fraction of interfaces. Since we measure that $P=1.22P_0$, we find that $c_i/c_w=(1.22-\phi _w)/\phi _i$. Using the values of $\phi _w$ and $\phi _i$ found using the confocal microscope images (0.25 and 0.12 respectively), one finds that $c_i/c_w \simeq 8$, showing that nile red is much more concentrated at the interfaces than in the continuous phase. Since the confocal microscope is spatially resolved, contrary to the spectrometer, voxels detecting interfacial regions receive more photons than voxels detecting regions of the aqueous phase, thus appearing brighter. To support this explanation, one can compute the ratio between the intensity received by this two kind of voxels (Fig. \ref{voxel}). One finds the value 8, in agreement with the above picture.\\
Having explained the method and its underlying effects, we now turn to some applications to illustrate the possibilities it raises. It has already been successfully applied to the study of forces distribution in such systems \cite{brujic}, and we want here to focus on applications based only on the localization of contacts.\\
%\vspace{-2cm}

\begin{center}
\begin{figure}[h!]
  \centerline{
    \mbox{\includegraphics[width=45mm]{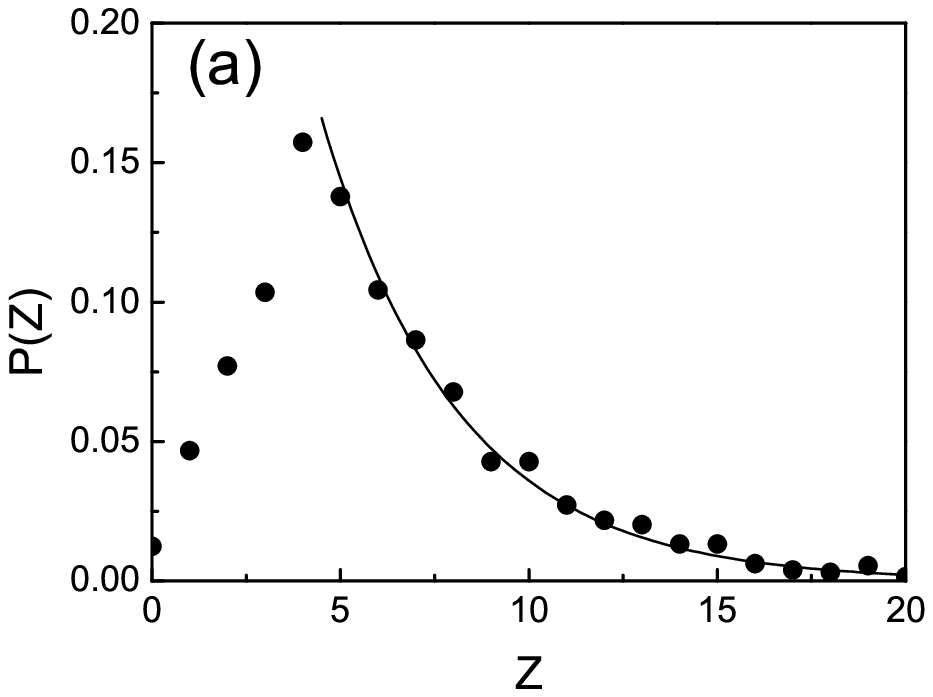}}
    \mbox{\includegraphics[width=45mm]{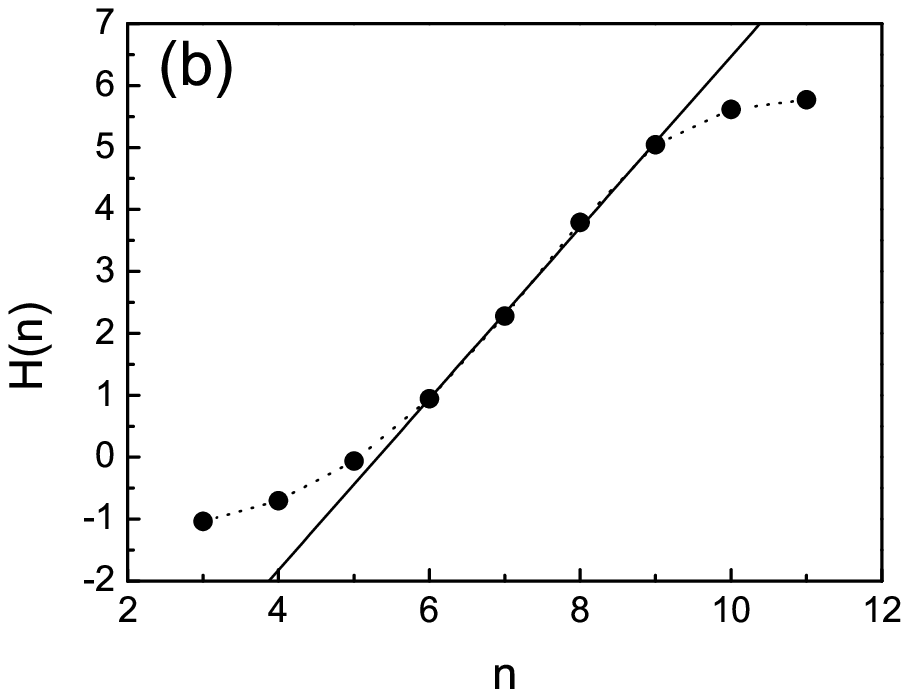}}
  }
\caption{(a) Distribution of the number of contacts per particle $Z$. The mean value is $\langle Z \rangle=6.08$. The solide line is an exponential fit of the tail. (b) $H$ as a function of the cluster size $n$.}
\label{fZ}
\end{figure}
\end{center}

{\it{Average number of contacts}} - The more trivial quantity which can be computed from the detection of contacts is the average number of contacts per particle. Roughly, with $N$ frictionless particles in dimension $D$, one has $ND$ Newton equations and $n_c/2$ unknowns, where $n_c$ is the total number of contacts, corresponding to the normal forces between droplets. Since $n_c=N\langle Z \rangle$, one finds that for the system to be solvable, it requires $\langle Z \rangle=2D$. In $3$ dimensions, this leads to $\langle Z \rangle=6$, the isostatic limit.\\
Experiments to measure this coordination number have been performed on grains. For instance, Bernal \cite{bernal} found $\langle Z \rangle=6$ (see also Donev \cite{donev}). This is surprising, since grains are frictional and the prediction in this case is $\langle Z \rangle=D+1=4$. Therefore, it is interesting to measure this quantity in a truly frictionless case.\\
The distribution of $Z$ is represented in Fig. \ref{fZ}a. The mean value extracted from this distribution is $\langle Z \rangle=6.08$, which is in remarkable agreement with the theoretical value of $6$. The slight discrepancy traducing an excess of contacts can be attributed to the light pressure exerted on the system. Note that the distribution P$(Z)$ exhibits an exponential tail due to the polydispersity of the particles which enables a more efficient packing, but despite this large polidispersity, the average value of $6$ still holds. This is an experimental proof of the prediction $\langle Z \rangle=2D$, for the case of frictionless particles at the colloidal lengthscale.
%\vspace{-1cm}

{\it{Entropy considerations}} - In the aim to enhance our understanding of the statistical physics of jamming systems, the ability to compute an entropy from microscopic variables is crucial. In the following, we show how it is possible to do such a task from the results obtained by the method presented above.\\
Determinig the entropy through the statistics of volumes, as suggested in numerous theoretical works \cite{ed3,blumenfeld} would require the knowledge of the density of states, which is still an open issue. We then use an alternative method which caracterizes the states of the system with graph theoretical representations and we compute the entropy density $s$ using information theory \cite{shannon,vink}.\\
To achieve this, we use our knowledge of the contacts positions to build the corresponding network. In this representation, a cluster of $n$ droplets is simply a graph which, by mean of a graph automorphism \cite{mckay}, can be transformed into a standard form (also known as "class") so that two graphs topologically equivalent belong to the same class. A given class $i$ can then be considered as a state with an occurence $p(i)$. In practice, one can determine $p(i)$ by extracting a large number $m$ of clusters of size $n$ from the system and counting the number of times $f_i$ a cluster $i$ is observed, so that:
\begin{equation}
\label{pi}
p(i)=\frac{f_i}{m}
\end{equation}
This allows to compute the "Shannon entropy" corresponding to clusters of size $n$:
\begin{equation}
\label{H}
H(n)=-\sum p(i) \ln p(i)
\end{equation}
where $i$ runs over all possible clusters of size $n$, and where we assume that the equivalent of the Boltzmann constant $\lambda=1$.\\
Using information theory, the entropy density $s$ is then calculated as follows:
%\vspace{0cm}
\begin{equation}
\label{ss}
s=\lim_{n \to \infty} [H(n+1)-H(n)]
\end{equation}
%\vspace{0cm}
With the values of $p(i)$ and equations (\ref{H}) and (\ref{ss}), it is then possible to determine $s$. Usually, it converges rapidly so that even moderate values of $n$ are enough to obtain a good approximation of $s$.
The result is shown in Fig. \ref{fZ}(right).
At large $n$, $H$ clearly suffers finite size effects, but the existence of a linear regime for intermediate values of $n$ demonstrates the rapid convergence of equation (\ref{ss}). Moreover, one notes that this linear regime, which demonstrates the extensivity of the entropy for cluster sizes starting from six grains, happens much faster than what has recently been observed in two-dimensional granular materials \cite{dauchot}.\\
The entropy per grain is then given by the slope of the linear part. One finds that $s \simeq 1.4$ (in units of $\lambda$). Beyond this number, the key point is that this method enables to clearly determine an entropy and then to study it as a function of the control parameters, such as the density of the system. This is crucial to investigate the relevance of the concept of entropy in the description of jamming and is the object of future work.\\
As a conclusion, we demonstrated a neat experimental technique for getting through a jammed system in a matter of seconds, giving direct and unambiguous access to contacts positions. This method has enabled us to perform an experimental proof of the average minimal coordination number $\langle Z \rangle$ for colloidal jammed systems and to investigate the concept of entropy in such systems. Our method thus allows the study of both quantities as a function of the density, thus reaching an equation of state for jammed systems. Furthermore, it is faster than X-ray tomography, thus permitting dynamical experiments with fast confocal microscopy. Many other theoretically daunting quantities can now be investigated in both static and dynamic experimental situations (shear bands, relaxation, etc). From micromechanics to statistical physics, this method promises to reveal new signatures of jamming.\\

Acknowledgements. We thank I. Hopkinson and H. Zhang for discussions.
We acknowledge support from DOE, Basic Energy Science,
Division of Materials Sciences and Engineering.

\end{document}